# Workload Engineering: Optimising WAN and DC Resources Through RL-based Workload Placement

Ruoyang Xiu
Google
rxiu@google.com

John Evans
Cisco
john.evans@cisco.com

## ABSTRACT

With the rise in data centre virtualization, there are increasing choices as to where to place workloads, be it in web applications, Enterprise IT or in Network Function Virtualisation. Workload placement approaches available today primarily focus on optimising the use of data centre resources. Given the significant forecasts for network traffic growth to/from data centres, effective management of both data centre resources and of the wide area networks resources that provide access to those data centres will become increasingly important.

In this paper, we present an architecture for workload placement, which uniquely employs a logically centralised controller that is both network and data centre aware, which aims to place workloads to optimise the use of both data centre and wide area network resources. We call this approach workload engineering.

We present the results of a simulation study, where we use a reinforcement-learning based placement algorithm on the controller. Results of the study show this algorithm was able to place workloads to make more efficient use of network and data centre resources and placed ~5-8% more workloads than other heuristic placement algorithms considered, for the same installed capacity.

## 1. BACKGROUND

With the rise in data centre (DC) virtualization, there are increasing choices as to where to place workloads, e.g. which may be represented by virtual machines (VMs) or containers. These choices may apply to web applications, to Enterprise IT and to Network Function Virtualisation (NFV), i.e. as defined by ETSI NFV [1]. Although the approach presented in this paper may apply to all cases, we focus on the NFV case.

The Cisco Visual Networking Index predicts that global Internet traffic will grow 3-fold from 2015 to 2020 [2]. Further, most Internet traffic today is originated or terminated in a DC and the Cisco Global Cloud Index predicts that cloud workloads will grow 3.3-fold from 2014-2019 and that global cloud IP traffic will account for more than 80% of total DC traffic by 2019 [3]. Hence, effective management of DC resources and of the networks resources that provide access to those DCs will become increasingly important.

Workload placement is a key resource management and scheduling function provided by DC virtualisation stacks (known as Virtualisation Infrastructure Managers, or VIMs in ETSI NFV terms) in use today [4], however, such solutions only consider the availability of resources within a DC {compute, memory, storage} when placing a workload. There are commercial solutions available today which augment such DC virtualisation stacks aiming to optimise workload placement by considering the availability of DC resources both within and between DCs [5], [6], [7], however, they do not consider availability of network resources. Mao et al [8] applied Reinforcement Learning (RL) to the problem of DC resource management, but similarly did not consider the availability of network resources.

Conversely, other works [9, 10] have proposed approaches to workload placement within a DC which aimed to optimize the use of network resources, however, they either did not consider availability of DC resources [9] or of network resources between DCs [9, 10]. Lin et at [11] applied RL to the problem of network resource management, but similarly did not consider the availability of DC resources.

In previous work [12], we proposed an approach to the placement of network traffic demands which aimed to optimize the use of IP network resources, i.e. which may provide wide area network (WAN) connectivity to and between DCs. This work did not consider availability of DC resources.

In practice, as workloads impact both DC and network resources, effective workload placement needs to consider both; there is no point in optimizing one domain, if the other is constraining. This balance will also change over time as demands grow and as network and DC resources are reprovisioned asynchronously. Further, with the rise of cloud-based services, effective placement often needs to consider choices between DCs as well as within DCs.

In this paper, we present an approach for workload placement which aims to optimize the use of both DC and WAN resources.

## 2. PROBLEM STATEMENT

We consider the placement of a workload which is defined by requirements for both DC and network resources and their associated service level agreements (SLAs); a

workload may be an aggregation of multiple DC resources and traffic demands:

- DC:
  - Resource requirements: the number of vCPUs required, memory required, storage required
  - SLA requirements: e.g. availability
  - Diversity requirements: affinity/anti-affinity to other workloads
  - Capability requirements: e.g. Single-Root Input/Output Virtualization (SRIOV), Data Plane Development Kit (DPDK)
- Network – per traffic demand:
  - Resource requirements: bandwidth
  - SLA requirements: latency, loss, availability
  - Diversity requirements: affinity/anti-affinity to other traffic demands

We consider a scenario with multiple candidate DCs where the workload may be placed; these DCs are interconnected by a WAN. Each DC may contain multiple clusters of host compute/storage resources, each of which is under the control of a local orchestration/control function; in NFV terms, Network Function Virtualisation Infrastructure (NFVI) instances under control of a VIM (e.g. Openstack). The scope of the problem is shown in Figure 1

**Figure 1. Workload Placement Scope**

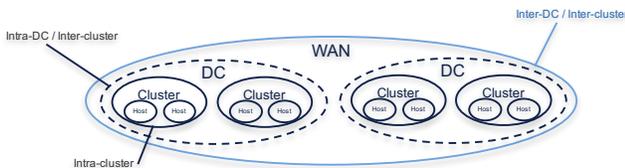

The primary focus of this paper is inter-DC, inter-cluster placement, as shown in Figure 2, which differentiates the scope of this paper from existing solutions and previous works.

**Figure 2. Scope for this paper**

|  | Intra-DC / Intra-cluster | Intra-DC / Inter cluster | Inter-DC / Inter-cluster |
|---|---|---|---|
| DC Resources | VIM [4], Mao [8] | Existing commercial solutions: [4], [5], [6] | This paper |
| Network Resources | Meng [9], Erickson [10] | | Lin [11], Evans [12] |

The workload will be accessed from multiple locations, which define one set of endpoints for the workload's traffic demands. The selected DC will define the other set of endpoints for the traffic demands.

The problem we try to answer is where can we place the workload to satisfying the following goals:

1. To support the workload requirements for both network and DC resources within the bounds of any defined SLAs?
2. To make most efficient use of DC and network resources?

## 3. WORKLOAD ENGINEERING ARCHITECTURE

To address inter-DC, inter-cluster placement, we assume the separation of global (intra-cluster) and local (inter-cluster) orchestration domains, which is a common deployment model. A single controller is used to make both global and local placement decisions; global placement determines which cluster to place the workload in and local placement determines where in that cluster to place the workload.

In this paper, we focus on the global placement problem and use the architecture shown in Figure 3 to address it. We note, however, that to be valid, the global placement decision must implement a superset of the local placement decision, sharing a common view of DC resources and employing the same local placement algorithm, from which it is able to derive the available DC capacity.

**Figure 3. Global Placement Architecture**

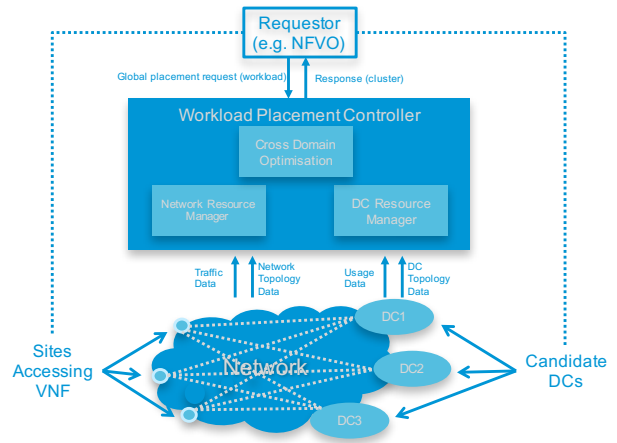

The architecture features a logically centralised workload placement controller; an overview of its operation is as follows:

1. Availability of DC resources is tracked by a DC resource manager function
2. Availability of network resources is tracked by a network resource manager function
3. Requestor. A requestor (e.g. NFVO) needs to create a new workload; the workload will require a set of DC resources and impose new bandwidth demands on the network. There are a set of potential DC locations where the workload may be sited $\{B_1, B_2, \ldots B_n\}$.
4. Workload Placement Controller API. Prior to creating the new workload, the requestor seeks guidance from a

controller via an API on which of the set of potential locations to use.

5. At a minimum, the request will include the following parameters:

   a. DC resources: the number of slots (S) required; we define a slot to be one quanta of the following resources required by a workload:
      i. Number of vCPUs required (V)
      ii. Memory required (M)
      iii. Storage required (H)
   b. Network resources:
      i. The A-end IP addresses for the traffic demands, i.e. which are fixed and may represent the set of endpoint locations that will be accessing the workload: $\{A_1, A_2, …, A_n\}$
      ii. The B-end IP addresses for the traffic demands, i.e. which represent the set of cluster locations where the workload may be sited: $\{B_1, B_2, … B_n\}$
      iii. The demand bandwidth required; this may be asymmetrical and may differ between traffic demand legs: $DA_1 \rightarrow B_x$, $DA_2 \rightarrow B_x$, … $DB_x \rightarrow A_1$, $DB_x \rightarrow A_2$, …
      iv. The maximum acceptable network latency: $L_{max}$

6. Workload Placement Controller. The controller is aware of the availability of network and DC resources and is able to determine what potential impact the workload will have on those resources and to determine which location is best, where best means is able to meet the SLA requirements and is optimal as determined by an optimisation policy.

   The logic applied by the controller is as follows:

   a. DC admission control: considering a specific cluster B:
      i. What is the minimum cluster capacity remaining in slots free (F) if the workload defined by S(V, M, H) is added?
      ii. F < 1: is an admission control failure, i.e. indicating there are insufficient DC resources in the cluster to support the workload. An admission control failure indication would be returned to the requestor.
      iii. F >= 1: there is sufficient capacity within the cluster ➔ consider network resources
   b. Network admission control: considering a specific A, B site pair:
      i. What is the minimum network capacity remaining (R) on all of the links on the paths between site $A_1$ and site $B_1$ if demand D is added?
         1. R < (1-T): is an admission control failure, which would indicate that there is insufficient bandwidth to support the requested SLA because the network utilisation exceeds the threshold of maximum acceptable bandwidth utilisation (T). T is set to ensure that the loss and jitter SLAs required for the workload can be met. An admission control failure indication would be returned to the requestor.
         2. R >= (1-T): there is sufficient bandwidth ➔ consider latency
      ii. What is the maximum latency ($L_{pmax}$) on all of the paths between site $A_1$ and site $B_1$?
         1. $L_{pmax} > L_{max}$: is an admission control failure, which would indicate the latency is too high to support the requested SLA because the path latency exceeds the maximum acceptable latency ($L_{max}$). An admission control failure indication would be returned to the requestor.
         2. $L_{pmax} <= L_{max}$: admission control success, i.e. indicating that the SLA can be met ➔ apply optimisation policy.

      This process is repeated for all A, B pair combinations; the set of combinations for which there has been an admission control success determines the list of feasible sites.

   c. Optimisation: if there are multiple feasible sites apply an optimisation policy to determine the preferred location.

7. Controller responds to requestor with selected DC
8. Requestor triggers instantiation of workload in selected DC
   a. The local orchestrator queries the workload placement controller for the required workload.
   b. The workload placement controller determines which host within the NFVI to place the workload in and responds to the local orchestrator
   c. The local orchestrator places the workload.

# 4. WORKLOAD ENGINEERING OPTIMISATION ALGORITHM

## 4.1 Reinforcement Learning and Workload Placement Problem

Meng [9] defines the traffic-aware VM placement problem and shows it to be NP-hard. Given the property of discrete and multi-metric balancing of the problem, it is natural to model it as a Markov Decision Process (MDP) and introduce Reinforcement Learning (RL) models to train an intelligent policy. RL employs agents that learn to make better decisions directly from experience interacting with the environment. The agent starts knowing nothing about the task at hand and learns by reinforcement — a reward that it receives based on how well it is doing on the task.

We use Q-learning, which is a model-free RL algorithm aiming to find an optimal action-selection policy for a given MDP. In our case, as the action-value function converges to the expected utility of taking a particular action in a given DC and network state, the agent would gain enough intelligence to make the best decision in each of the different conditions.

We create a history by training q-learning with a series of workloads with different requirements. The resulting action-value function is the intelligent core of the system.

## 4.2 Q-Learning Based Workload Placement Model

An initial Q-Learning model is proposed to fit the workload placement problem into a MDP as shown in Figure 4.

**Figure 4. RL Model**

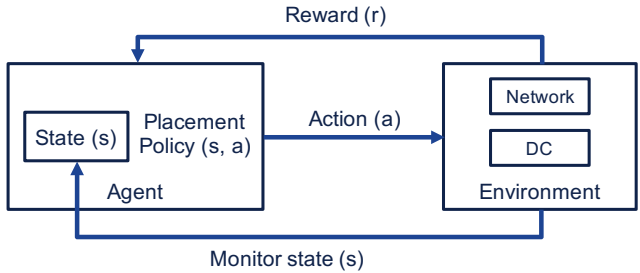

The tuples of the MDP are defined as follows:

- *State (s)*. Three factors are used to fully describe the current condition: the incoming workload requirements, the current utilisations of all network paths and the current availability of DC resources. We introduced the concept of slots to match DC resources with the workload to abstract the discrete workload requirements and to constrain the cost of depicting the topology (i.e. a connected graph).

  The initial state at time unit t is:
  - A vector of the maximum path utilisation of all network paths from a DC to all of its access points
  - A vector of DC resources in terms of slots free of all DCs.

- *Action (a)*. The action set is a selection of policies that the agent can choose from to place an incoming workload. We define three optimisation policies as shown in Table 1. The RL agent learns to pick the one among the three policies per workload to gain most rewards in the long-term. The way the action set is defined avoided exponential gain when dealing with complex topologies, whilst maintaining diversity (in most cases the three pre-defined policies would result in a different placement decision).

**Table 1. RL optimization factors**

| Policy Name | Description |
|---|---|
| DataCentreOpt | Choose the DC/VIM which has the maximum available DC capacity in slots free, i.e. to optimise use of DC resources |
| PathUtilOpt | Choose the DC with the lowest network path utilisation across the constituent demands of the workload, i.e. to optimise use of network resources |
| LatencyOpt | Choose the DC with the lowest average network path latency across the constituent demands of the workload |

- *Reward (r)*. Reward directly determines the goal of the model. The rewards are defined to keep the maximum path utilisation as low as possible and the available DC capacity as high as possible:

$$reward_t = \begin{cases} score(p_t) + score(d_t), if\ placed \\ -1000, otherwise \end{cases}$$

Where $p_t$ is *(1 – max-path-util%)* and $d_t$ is the *(slots-free%)* after placing workload at time t, where the maximum value of each is 1.

We have two termination conditions: when either the maximum path utilisation reaches 100% or there is no available DC capacity for the current workload, the simulation will stop.

# 5. SIMULATION STUDY

We undertook an ad-hoc simulation study focussed on workload placement, using real network topologies from two network service provider to determine the potential performance of workload engineering using Q-learning compared to other workload placement algorithms.

One of the network topologies was from an international service provider interconnecting 11 access points of presence (POP) locations, 7 of which provided connectivity to DC resources; the network consisted of 10Gbps links. The other was from an in-country service provider interconnecting 5 access POP locations, 2 of which provided connectivity to DC resources; the network consisted of 80Gbps links. We omit the actual topologies for reasons of confidentiality.

100 simulation iterations were run for each algorithm. The same workload sets were used for all algorithms – with over 1,000 randomly generated workloads for each iteration:

- DC. Randomly selected DC requirements per workload with a uniform distribution of each metric:
  - 2, 4, 8 vCPUs
  - 4GB, 8GB, 16GB RAM
  - 256GB, 512GB, 1024GB storage
- Network:
  - Each workload is accessed from a uniformly random selection of the access POPs
  - The constituent demands for each workload were sized symmetrically and uniformly across the workload, i.e. each demand was the same bandwidth as the others in the workload
  - The individual demand bandwidth was randomly selected for each workload from: 128Mbps, 256Mbps, 512Mbps

In each case we discount any DCs/VIMs that have insufficient DC or network resources to support the workload and then apply an algorithm to select between the feasible DCs. The following workload placement algorithms were compared:

- *Random* – for each workload the serving DC is selected randomly. This is considered a baseline.
- *DataCentreOpt* – to optimise use of DC resources, as per Table 1.
- *PathUtilOpt* – to optimise use of network resources as per Table 1.
- *LatencyOpt* – to minimise network latency as per Table 1.
- *Q-learning* – RL-based placement algorithm, which aims to maximise the number of workloads placed by choosing between these policies depending up the state: *DataCentreOpt, PathUtilOpt, LatencyOpt* as per Table 1.

When there are no feasible DCs capable of supporting the requested resources, i.e. because there are either not enough network or DC resources available, the simulation stops.

## 5.1 Training

The action matrix was trained using 100,000 workloads – defined as described above. When there were no feasible DCs to support a workload, a large negative reward was incurred, the state map was cleared and the training continued. Figure 5 shows the total reward gain during training per 1000 placements.

**Figure 5. Total reward gain per 1000 placements**

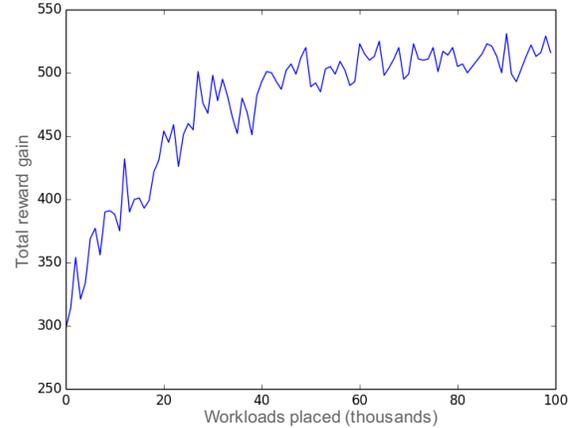

## 5.2 Simulation Results

We compare the algorithms in terms of numbers of workloads successfully placed, i.e. the more workloads placed the more effective the algorithm.

Figure 6 shows the aggregate results comparing the average number of workloads placed for the five workload placement algorithms across all 100 simulation runs, normalised by the random results; the error bars indicate the minimum and maximum percentage of workloads placed:

**Figure 6. Aggregate Results**

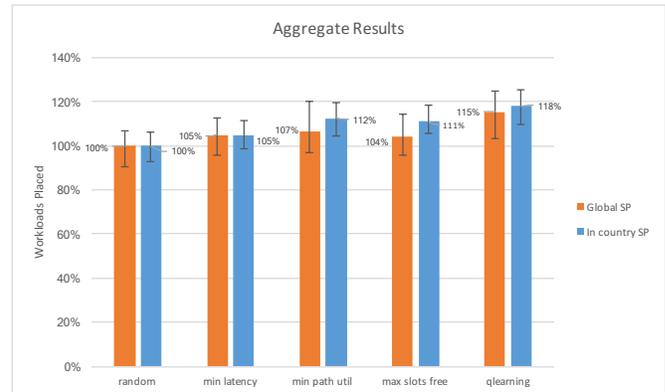

The achievable benefit for any scenario varies with resource availability and demands; this varies by workload and topology. Q-learning outperformed random placement, placing on average ~15% more workloads for the global SP, and ~18% more workloads for the in-country SP. Q-learning outperformed the other heuristic placement algorithms, placing on average ~8% more workloads than the next best performing algorithm for the global SP, and ~5% more workloads for the in-country SP.

Considering the individual simulation runs, Q-learning outperformed the other placement algorithms, adapting to the available resources and placing more workloads in 89% of the simulation runs for the global SP and 91% of the simulation runs for the In-country SP, as shown in Table 2:

Table 2. % of simulation runs for which each algorithm placed the most workloads

|  | In country SP | Global SP |
| --- | --- | --- |
| Random | 0% | 0% |
| Minimise Latency | 0% | 2% |
| Minimise Path Util | 8% | 7% |
| Maximise Slots Free | 2% | 2% |
| Q-learning | 91% | 89% |

## 6. FURTHER STUDY

We highlight the following as potential areas of further study:

- We focus on inter-DC, inter-cluster placement in this paper. The greatest potential gains stand to be attained when inter-DC, inter-cluster placement is combined with intra-cluster placement.

- In this study we simplified the DC resource management parameters by abstracting them to slots to limit the cost of depicting the topology. Exposing the workload details and applying deep reinforcement learning may provide further adaptability and optimisation.

## 7. CONCLUSIONS

In this paper we presented an approach to workload optimisation leveraging a centralised controller that is aware of both network and DC resources; we call this approach workload engineering. We employed an RL-based placement algorithm on that controller which was able to adapt to the available resources and make efficient choices of where to place the workloads, placing 15-18% more workloads than random placement and outperforming the other heuristic placement algorithms considered, placing 5-8% more workloads for the same installed capacity.

As part of this study, we also built a working implementation of workload engineering [13].

## 8. ACKNOWLEDGEMENTS

We would like to thank Professor Albert Cabellos and Nick Hall for their useful feedback and insights in producing this paper. We would also like to thank Baton Daullxhi, for his part in building the working implementation.